\documentclass[epj,referee]{svjour}

\newcommand {\be} {\begin{equation}}
\newcommand {\ee} {\end{equation}}
\newcommand {\Be}{\begin{eqnarray*}}
\newcommand {\Ee} {\end{eqnarray*}}
\newcommand {\bey} {\begin{eqnarray}}
\newcommand {\eey} {\end{eqnarray}}

\usepackage{graphics}
\begin{document}

 \title{ Memory Effects and Heat Transport in 
One-dimensional Insulators}
 
\author{Stefano Lepri \inst{1} \inst{2}
\thanks{E-mail: {\rm lepri@avanzi.de.unifi.it}}
}
\institute{Dipartimento di Energetica ``S. Stecco'', 
Via Santa Marta 3 I-50139, Firenze (Italy) \and
Istituto Nazionale di Fisica della 
Materia, Unit\`a di Firenze, L.go E. Fermi 2 I-50125, Firenze (Italy)}

\date{\today}
\abstract{
We study the dynamical correlation 
functions and heat conduction for the simplest model of quasi one-dimensional
(1d) dielectric crystal i.e. a chain of classical particles coupled by 
quadratic and cubic intersite potential. Even in the weakly nonlinear regime,
numerical simulation on long enough chains reveal sizeable deviations from 
the perturbative results in the form of a slower decay of correlations in
equilibrium. Their origin can be traced back to the subtle nonlinear effects
described by mode-coupling theories.  
Measures of thermal conductivity with nonequilibrium molecular-dynamics 
method confirm the relevance of such effects for low-dimensional 
lattices even at very low temperatures. 
}
\maketitle

\section{Introduction}

The theoretical study of physical models in one spatial dimension is
often justified by their mathematical simplicity \cite{mattis}. 
Besides this, they display intriguing peculiarities that render
them qualitatively different from their three-dimensional counterparts.
Actually, a further and even more relevant motivation is the possibility
of modern experimental techniques to produce a variety of real systems
that could , at least in principle, be effectively described by 1d models. 
Furthermore, the latter are of great importance to approach
the dynamical and statistical properties of biological molecules like 
DNA \cite{dauxois} or proteins \cite{stillinger}.   

In the present paper we wish to study the problem of relaxation and heat
transport in a classical chain of oscillators with quadratic and cubic 
intersite potential. Although this is a truly textbook model of an 
insulating crystal, there is no, to the best of our knowledge, systematic 
analysis of the effects of anharmonicity. The results may be of interest
to describe experimental systems like e.g. strongly anisotropic  crystals 
\cite{morelli,smontara}, polymers \cite{forsman} or nanowires \cite{tighe}.
Some theoretical investigations of thermal conductance for a quantum wire in 
ballistic \cite{rego} and anharmonic \cite{leitner} regimes have been 
also  recently presented.
 
The first issue that we address is the following: which difference should 
one observe in this case with respect to the usual 3d systems? 
One may argue that the cubic nonlinearity (``three phonon" scattering)
must be less efficient due to the stronger constraints to be fulfilled in 
1d, an argument that has been even invoked to explain several
experimental results \cite{morelli,smontara}. A signature of this fact
on the dynamical correlations is thus to be expected. To substantiate 
this argument, we studied the model within the framework of the memory-function 
formalism \cite{kubo}. A perturbative calculation shows that correlation 
functions display a power-law tail (Section II), that we study in some
detail in the long-wavelength limit and compare, to some extent, with 
molecular dynamics simulations. The latter show that, even for relatively
small anharmonicity, strong non-perturbative effects due to subtle nonlinear
interaction arise for long enough chains. These can be quantitatively
accounted for by mode-coupling theories \cite{lepri} akin to the ones usually
applied for dense fluids \cite{pomeau}. 

The consequences of such a behaviour on nonequilibrium energy transport
is discussed in Section III. Even in absence of such strong memory
effects, the thermal conductivity has been shown to diverge with the 
system size both for 1d \cite{noi,noi2} and 2d lattices \cite{lippi}. 
The characteristic divergence law, a signature of the mentioned 
mode-coupling effects \cite{noi2}, is recovered also in the present case. 
This signals again the overwhelming role of nonlinearity and 
fluctuations in low spatial dimensions.  
 
\section{Correlation functions for an anharmonic chain}

We consider a set of atoms of mass $m$ arranged on a ring of $N$ sites with 
spacing $a$. Let $u_l$ be the displacement of the $l$-th particle from its 
equilibrium position $la$. The Lagrangian reads as  
\be
{\cal L} = \sum_{l=1}^N \left[{m\over 2} \dot u_l^2-
{m \omega_0^2 \over 2} \, (u_{l+1}-u_{l})^2
-{g\over 3}\,(u_{l+1}-u_{l})^3  \right] 
\quad;
\label{lagra}
\ee
where $g$ is the coupling constant. This is the 
lowest-order approximation of a generic anharmonic potential which couples
nearest neighbours. For historical reasons, it is 
sometimes referred to as the Fermi-Pasta-Ulam $\alpha$-model
(FPU-$\alpha$). For convenience, 
from now on we will always work in dimensionless units, where $a$, $m$ and the 
angular frequency $\omega_0$ are set to unity. This implies, for instance,
that the sound speed $a\omega_0$ is also unity and  
that the energy and coupling constant (that we still denote with $g$)  
are measured respectively in units of $m\omega_0 a^2$ and $m\omega_0 a^{-1}$.

A difficulty of model (\ref{lagra}) is the unboundedness of the 
potential. Therefore, in order to avoid runaway instability of
trajectories (which could anyhow be overcome by adding even terms of higher
order) we will deal with sufficiently small coupling constant and/or energies.
 
Upon introducing the complex amplitudes $U_k$ through the 
discrete Fourier transform  
\be
U_k={1\over\sqrt{N}}\sum_{l=1}^N u_l \, e^{i{2\pi k\over N}l} \quad, 
\label{modi}
\ee
where $k$ is an integer ranging between $-{N\over2}+1$ and ${N\over2}$
we can thus rewrite the equations of motion as
\be
\ddot U_k +\omega_k^2 U_k \;=\;
-g \omega_{k}\,{1\over \sqrt{N}}\sum_{k_1+k_2=k}
\; \omega_{k_1}\omega_{k_2} U_{k_1}U_{k_2} \;\equiv\;
{\cal F}_k
\label{newton}
\ee
with ${\cal F}_k$ being the interaction force among modes. 
The condition on the indices of the sum is intended to be 
modulo $N$ while 
\be
\omega_k \;=\; 2 \, \big| \sin\left({\pi k\over N}\right)\big| \quad.
\label{barefreq}
\ee
is the usual harmonic (phononic) dispersion law. 

\subsection{Memory effects}

Let us consider the normalized correlation function
\be
{\cal G}_k(t) \;=\; \beta \omega_k^2\, \langle U_k(t)U_{k}^*(0)\rangle \quad,
\ee
which is defined in such a way that ${\cal G}_k(0)=1$ ($\beta$ is the inverse 
temperature). In the framework of the Mori-Zwanzig projection approach, it 
can be shown that it satisfies the equation of motion \cite{kubo}
\be
\ddot {\cal G}_k 
+\int_0 ^t \Gamma_k(t-s) \, \dot {\cal G}_k(s) ds 
+ \omega_k^2 {\cal G}_k \; = \; 0 \quad .
\label{eqcor}
\ee
The memory function $\Gamma_k$  accounts for  memory effects and can 
be connected to the nonlinear force by the 
fluctuation-dissipation relation
\be
\Gamma_k(t) \;=\; \beta 
\langle {\cal F}_k(t){\cal F}_k^*(0)\rangle \quad.
\label{fludis}
\ee
Notice that here it is also implicitely assumed that slow terms possibly 
contained in ${\cal F}_k$ are negligible in the thermodynamic limit 
\cite{lepri}.

In the following we will evaluate the memory function to the lowest order
of perturbation theory. This amounts to evaluating the average in (\ref{fludis})
on the measure of the unperturbed system ($g=0$). A straightforward calculation 
yields 
\be
\Gamma_k(t) \;\approx\;
{C g^2\omega_k^2 \over \beta}\, {1\over N} \sum_{k_1+k_2= k}
\cos \omega_{k_1}t \; \cos \omega_{k_2}t \; \quad
\label{corfo3}
\ee
where $C$ is a suitable numerical constant.
Rather remarkably, the Laplace transform of $\Gamma_k$ can be 
evaluated exactly in the large $N$ limit (see the Appendix for some details 
of the calculation). Indeed, if we let $2\pi k/N \to q$ and define 
\be
\Gamma(q,z)\;=\;\int_0^\infty  \, \Gamma(q,t) e^{-izt} \, dt 
\ee
we obtain the simple result
\be
\Gamma(q,z) = K \omega^2(q) \left[ {1\over \sqrt{\Omega_+^2(q) -z^2}}
+{1\over \sqrt{\Omega_-^2(q) -z^2}} \right] \quad ,
\label{beslap}
\ee
where we have introduced the new (small) coupling parameter 
$K=C g^2 / 2 \beta$ and 
\be
\Omega_+(q)= 4\cos {q \over 4} \quad,\quad 
\Omega_-(q)= 4 |\sin {q \over 4}| \quad.
\label{omegapm}
\ee
Inversion of the Laplace transform yields \cite{abram}
\be
\Gamma(q,t) \;=\; K\omega^2(q) \,\left[ J_0(\Omega_+(q)t) + J_0(\Omega_-(q)t)
\right] \quad ,
\label{bessel}
\ee
where $J_0$ is the Bessel function. With this result at hand, we can give 
a closed expression for ${\cal G}$ by solving Eq. (\ref{eqcor}) with the 
Laplace transform method that yields the formal solution 
(with $\dot{\cal G}(q,t=0)=0$)
\be
{\cal G}(q,z) \;=\; 
{iz+\Gamma(q,z) \over z^2 - \omega^2(q) -iz \Gamma(q,z)} \quad .
\label{laplag}
\ee
If, like in the present case, the dissipative effects are weak, the latter 
expression can be approximated as
\be 
{\cal G}(q,z) \;=\;
{-i/2\over z-\omega(q) - i\Gamma(q,z)/2} +  
{-i/2\over z+\omega(q) - i\Gamma(q,z)/2}
\label{approxg}
\ee
and the correlation function are determined for every $q$ by the standard 
methods of inverse Laplace transformation i.e. by determining the 
singularities in the complex plane of ${\cal G}$.

\subsection{Long wavelengths}

Let us focus on the effective dynamics of the long-wavelength modes.
The latter are the slowest ones and are directly responsible of energy
transport and therefore of main interest here.
For $q\to 0$ one has from Eq. (\ref{omegapm}) $\Omega_+\approx 4$ 
and $\omega \approx \Omega_-\approx  |q| $. This implies that the
first term of the memory function decays on a much faster time scale and
can be neglected so that
\be
\Gamma(q,z) \;\approx\; {Kq^2\over\sqrt{q^2-z^2}} \quad .
\label{smallq}
\ee
Therefore, in this limit, the transform of the mode autocorrelation 
satisfies the scaling relation 
\be 
{\cal G}(q,z) \;=\; {1\over|q|}G\left({z\over |q|}\right) \quad,
\label{scaling}
\ee
where, from Eq. (\ref{approxg})
\be
G(w)\;=\; {-i/2\over w-1 - iK/2\sqrt{1-w^2}} +  
{-i/2\over w+1 - iK/2\sqrt{1-w^2}}
\label{Gw}
\quad .
\ee
From Eqs.(\ref{scaling}) and (\ref{smallq}) it follows immediately 
that ${\cal G}(q,t)$ depends only on the product $|q|t$. 

To estimate the behaviour of ${\cal G}(q,t)$ in a more detailed way
one has to study the properties of the function $G(w)$. Besides of the branch 
cut in $z=\pm 1$, the latter has two simple poles in the upper part of the 
complex plane that, for small $K$, are approximatively given by
$\pm 1 \, + \, i{\sqrt{3} \over 4}K^{2/3}$ (we neglect the 
term ${\cal O}(K^{2/3})$ in the real part since 
this gives only a small correction to the unperturbed frequency).
Accordingly, the inversion of the Laplace transform (\ref{Gw})
yields two contributions to the decay, the exponential plus some power-law 
tails. More precisely, the calculation yields
\be
{\cal G}(q,t) \;\propto\; \exp(- \gamma (q)|t|) \,
\cos qt \,  - K f(K,|q|t)
\label{correlaz}
\ee
where we have introduced the (short times) relaxation rate
\be 
\gamma(q) \;=\;  {\sqrt{3} \over 4}K^{2/3}\,|q| \quad .
\ee
Notice how the square-root in (\ref{Gw}) affects the functional
dependence of the relaxation time on the coupling constant and therefore 
on temperature. The behaviour of the slowly decaying correction
\be
f(K,\tau) \;=\; {1\over 2\pi} \int_{-1}^{+1} \, dw
{\sqrt{1-w^2} \over (1-w^2)(w+1)^2+K^2/4} \, \cos w\tau \quad .
\label{effe}
\ee
has been evaluated numerically finding that, as expected, the
latter oscillates with a period close to $2\pi$ while its
envelope decays for large $\tau$ as $K^{-2}\tau^{-3/2}$.

\subsection{Comparison with the numerical simulations}

In the present section we compare the results of the perturbative calculation
with the outcomes of molecular dynamics simulations. The latter were performed
at equilibrium and in the microcanonical ensemble by integrating the 
equations of motion with a third order symplectic algorithm \cite{algor}
and starting from random initial conditions. We then let the system evolve 
for a certain transient time in order to start the measures from a generic 
phase-space point. The energy per particle has been fixed in every computation 
and the corresponding temperature has been measured as twice the average 
kinetic energy density. In computing spectra 
and correlations functions a Fast Fourier Transform routine has been used, and 
the data are usually averaged over an ensemble of several trajectories 
(typically between 20 and 200) to reduce statistical fluctuations.

\begin{figure}[h]
\resizebox{0.45\textwidth}{!}{\includegraphics{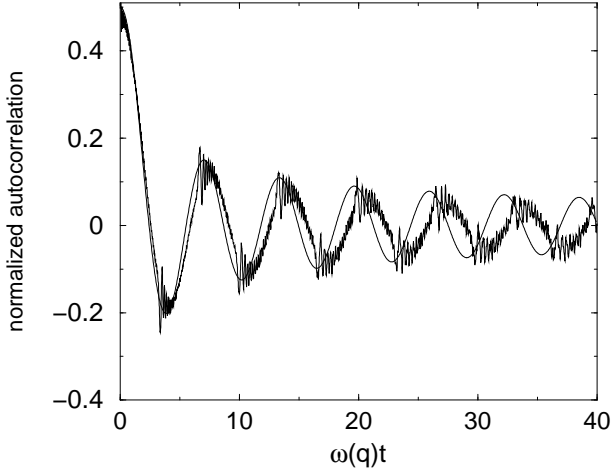}}
\noindent
\caption{The normalized autocorrelation function of the 
nonlinear force ${\cal F}_k$ for $\beta =10^{4}$, 
$g=1.0$, $N=128$, $k=1$ (corresponding to $q=0.0490$). The thick 
solid line is the (approximate) 
perturbative result $J_0/2$ (see eq.(\protect\ref{bessel})).}
\label{corf}
\end{figure}

\begin{figure}[h]
\resizebox{0.45\textwidth}{!}{\includegraphics{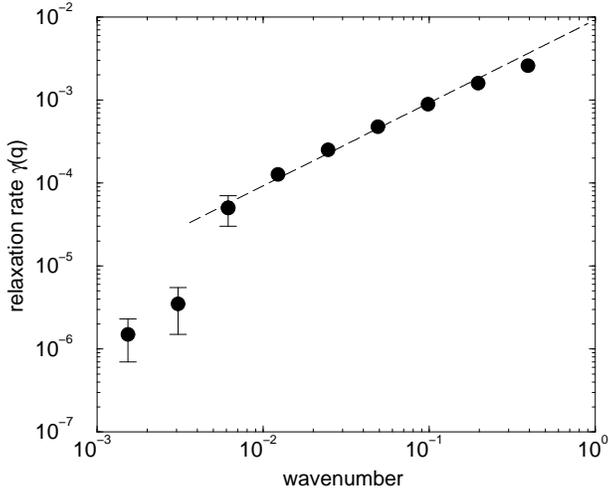}}
\noindent
\caption{
Scaling of the relaxation rates with the wavenumber $q$ for 
$g=0.25$ and $\beta=10.0$. The straight line is the result 
of perturbation theory.
}
\label{ratesq}
\end{figure}

\begin{figure}[h]
\resizebox{0.45\textwidth}{!}{\includegraphics{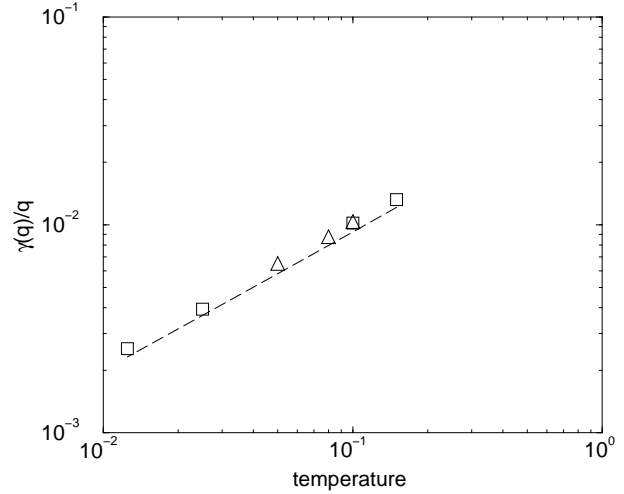}}
\noindent
\caption{
Scaling of the relaxation rates with temperature
for the first Fourier mode of a chain of length $N=256$ (triangles)
and $N=512$ (squares). The straight line is the result 
of perturbation theory.
}
\label{ratese}
\end{figure}

As a first check we compared the perturbative expression of the memory
function (\ref{bessel}) with the correlation of the nonlinear force
appearing in Eq. (\ref{fludis}) (see Fig. \ref{corf}). 
The two agree well on short times, but some deviation is observed after
a few oscillation periods.
Furthermore, the temperature and wavenumber dependence of the relaxation 
rates has been checked by measuring the initial decay rate of the 
envelope of ${\cal G}(q,t)$. As seen in Figs. \ref{ratesq} and \ref{ratese}, 
a reasonable agreement with the result of 
perturbation theory is obtained for $q >0.006$ (corresponding to $N<10^3$).
However, for smaller wavenumbers (longer chains) the data crossover to 
a different behaviour that is consistent with the law $q^{5/3}$. The
latter is the result expected from mode-coupling theory and has been 
numerically observed for the model with quartic nonlinearity at much 
larger temperatures \cite{lepri}.

We can therefore conclude that, despite the smallness of the perturbative 
parameter ($K\sim 0.01$), the dynamical effects of nonlinear mode 
interaction take over on longer (``hydrodynamic") time/length scales. 
Eventually, this originates substantial deviations from the results 
obtained in the previous section.

\section{Thermal conduction}

In the present Section we will study the consequences on the phenomenon of 
stationary heat transport along the chain. 
Let us discuss the case of linear response i.e. small thermal gradients.
We first evaluate the asymptotic decay of the Green-Kubo integrand
in the framework of the perturbative calculation described above and compare
the results with numerical simulations.

\subsection{The Green-Kubo formula}

If we neglect the anharmonic term (which contributes to the 
conductivity only to order $g^2$), the heat current $J$ is approximated
by its harmonic part \cite{noi2}
\be
J\;\approx\; J_H = {i\over 2N} \sum_{k=-N/2 +1}^{N/2}
\, v_k \omega_k \left(U_k\dot U_k^* -  U_k^*\dot U_k\right)
\label{armoflux}
\ee
where $v_k=\omega_k'$ is the phase velocity. It is convenient to recast
the above expression in the new variables $A_k=e^{i\omega_k t}U_k$.
Taking into account the fact that $\sum_k v_k \omega_k |A_k|^2=0$ for
symmetry reasons ($v_k=-v_{-k}$), one finds
\be 
J_H = {i\over 2N} \sum_k v_k \omega_k
\left(A_k\dot A_k^* -  A_k^*\dot A_k\right) \quad.
\ee 
The integrand appearing in the Green-Kubo formula for thermal conductivity
\be
\kappa \;=\; k_B\beta^2 N \int_0^{\infty} \langle  J(t)J(0) \rangle dt
\quad .
\label{gkubo} 
\ee
contains correlation functions of quantities that are quadratic in $A_k$.
As the latter quantities are slowly varying ($\dot A_k \ll \omega_k A_k$)
we get the approximate expression  
\be
 N \langle  J(t)J(0) \rangle \;\approx\;
 {1\over 4\beta N} \sum_k v_k^2
\, \langle \dot A_k(t) \dot A^*_k(0) \rangle
\,+ \, c.c.  
\label{kuboint}
\ee
where we have also taken into account the equipartition of energy 
$\langle \omega_k^2 |A_k|^2 \rangle = 1/\beta$.

Let us assume that the large $t$ behaviour of (\ref{kuboint}) is
dominated by long-wavelength modes. On the basis of the above discussion
one expects that in this limit the autocorrelation of $\dot A(q,t)$
will depend only on $|q|t$, i.e.   
\be
\langle \dot A(q,t) \dot A^*(q,0)\rangle \;=\; {1\over \beta} 
{\cal A}(|q|t)\quad, 
\ee
where ${\cal A}$ is a suitable function. Replacing the sum in 
(\ref{kuboint}) with an integral and taking into account the fact that 
$v(q)\to 1$ for $q\to 0$ one gets
\be 
N \langle  J(t)J(0) \rangle \; \propto \; {1\over t \beta^2}
\int_0^{+\infty} {\cal A}(x) dx
\ee
Since that the integral in this formula is convergent because of the
asymptotic behaviour of (\ref{effe}), we can conclude that the Green-Kubo
integrand decays as $1/t$ for large $t$ and that the thermal conductivity
is infinite. 

To estimate the divergence law for a finite chain, one can cut off the 
integral in the formula (\ref{gkubo}) up to some time proportional to the 
length $N$ \cite{noi2}. As a result, $\kappa$ should diverge logarithmically
with the system size. In the next subsection we will compare this prediction 
with numerical results.

\subsection{Nonequilibrium simulations and finite-size conductivity}

A simple and efficient way to compute transport coefficients is to use
nonequilibrium molecular dynamics. First of all we consider the chain
with fixed ends ($u_0=u_{N+1}=0$) and let the first and the $N$-th oscillators 
interact with two reservoirs operating at different temperatures 
$T_R=T+\Delta T/2$ and $T_L=T-\Delta T/2$, respectively. In such a way a 
net heat current flows through the lattice. The thermal conductivity for the 
finite chain can thus be measured as the ratio between such a flux and the 
applied temperature gradient.

Among the several possible choices, we simulated the effect
of the reservoirs by means of Nos\'e-Hoover thermostatting method \cite{nose}. 
The latter preserves the deterministic nature of the dynamics and is simply
implemented by adding the force terms $-\zeta_L\, \dot u_1$ and 
$-\zeta_R\, \dot u_N$ to the equation of motion of the first and last
oscillator respectively. The ``thermal'' variables $\zeta_L$, $\zeta_R$, 
evolve according to the dynamical equations \cite{nose}
\bey
\label{nosehoover}
&&\dot \zeta_L \;=\; {1\over \tau^2}\left({{\dot u_1}^2 \over T_L}
  -1\right) \nonumber\\
&&\dot \zeta_R \;=\; {1\over \tau^2}\left({{\dot u_N}^2 \over T_R}
  -1\right) \quad ,
\eey
where $\tau$ is the thermostat response time and controls
the strength of the coupling between the reservoirs and the chain.
The above prescriptions imply that the kinetic energy of the boundary
particles fluctuates around the imposed average value, thus mimicking
an interaction with a reservoir in canonical equilibrium. 

The simulations were performed integrating the equation of motion for the
bulk particles together with the (\ref{nosehoover}) with a fourth-order
Runge-Kutta algorithm. The boundary temperatures were kept fixed so that 
upon increasing the lattice length the applied temperature gradient 
$\Delta T/N$ is decreased and the Green-Kubo 
formula becomes more and more accurate. Both the time averaged kinetic 
temperatures $T_l=  \overline{\dot u_l^2} $ and  heat flux $\overline{J}$ 
where \cite{choq,noi3}
\begin{equation}
J  ={1\over 2N} \sum_l (\dot u_{l+1}+\dot u_l)\, 
\left[u_{l+1}-u_l + g(u_{l+1}-u_l)^2\right]   ,
\label{fluxtot}
\end{equation}
have been computed over a single trajectory (approximatively $10^6$ time 
units) started from random initial conditions . In every run, 
a suitably long transient (about $10^4$ time units) has been discarded in order
to let the system reach a statistically stationary state with each oscillator 
in local equilibrium.  

The thermal conductivity is thus computed as $\kappa = |\overline{J }| N/\Delta
T$. It has to be noted that the latter quantity represents an effective
transport  coefficient including both boundary and bulk scattering mechanisms. 
Alternatively, one could as well consider, say, a subchain far enough from the
ends and compute a bulk conductivity as the ratio between $|\overline{J}|$ and
the actual temperature gradient measured there. Since the latter will also be
inversely proportional to $N$, the resulting scaling with size (which is the
main issue here) will be of course the same with both definitions. Besides 
this, the first choice avoids the difficulty of dealing with very small 
gradients (see below). 

\begin{figure}[h]
\resizebox{0.45\textwidth}{!}{\includegraphics{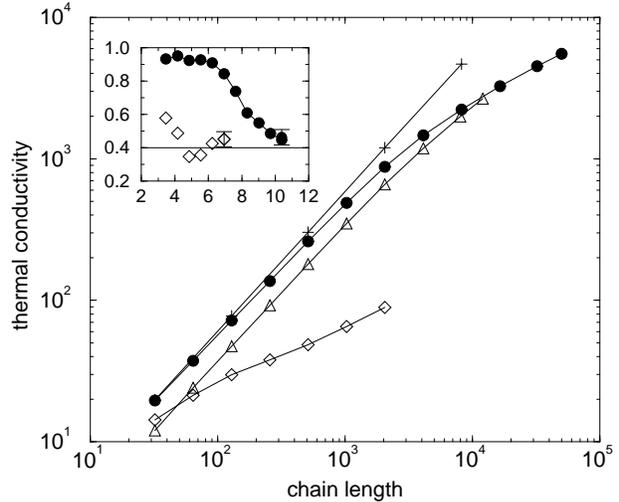}}
\noindent
\caption{
Thermal conductivity of the FPU-$\alpha$ model versus lattice length 
$N$ for $g=0.25$, $T=0.1$, 
$\tau=1.0$ and $\Delta T=0.1$ (triangles), $\Delta T=0.02$ (full circles). 
The crosses (resp. diamonds) refer to the harmonic (resp. FPU-$\beta$) models 
for $T=0.1$ and $\Delta T=0.02$. The inset shows the logarithmic derivative 
$\Delta\log \kappa/\Delta\log N$ versus $\log N$ of the data for the 
FPU-$\alpha$ (full circles) and FPU-$\beta$ (diamonds) respectively.
}
\label{conduct}
\end{figure}

The dependence of $\kappa$ on the chain length is reported in Fig. \ref{conduct} 
for two different series of simulations with $T=0.1$ and for relatively small 
($\Delta T=0.02$) and large ($\Delta T=0.1$) applied temperature differences.
In both cases $\kappa$ increases linearly with $N$ for $N< 10^3$. 
Moreover, no sizable temperature gradient forms along the chain. Both
facts are a signature of the weakness of the anharmonic effects up to this 
time/length scales, as confirmed by comparing  with data 
obtained (using the same setup and parameters) for the pure harmonic ($g=0$) 
chain. The latter displays the expected linear increase of $\kappa$ with $N$ 
\cite{lieb} and differ by less than a few percent from the anharmonic case 
in this range of system sizes (compare full dots and crosses in 
Fig. \ref{conduct}). The fact that $\kappa$ is smaller for larger $\Delta T$ 
can be thus be attributed to stronger boundary scattering that, in turn,  
tends to reduce the conductivity. 

For larger $N$ the bulk anharmonic scattering takes over and, accordingly, 
the two curves for different $\Delta T$ approach one the other. The 
conductivity increases more slowly with the size, and we can tentatively 
measure the effective divergence law of the form $N^\alpha$. The data in
the inset of Fig. \ref{conduct} (full dots) are consistent with a convergence
to the asympotic value $\alpha=2/5$ expected from the 
theory and observed in previous works on other 1d models \cite{noi2,hatano}. 
Notice also that the crossover size is in reasonable 
agreement with the behavior reported in Fig. \ref{ratesq}.

In order to clarify the role of different anharmonic interactions,  we also
compared the above results with some measurements for a chain with quartic
potential. More precisely, we considered the so-called  Fermi-Pasta-Ulam
$\beta-$model (FPU-$\beta$)  where the cubic term in the lagrangian 
(\ref{lagra}) is replaced by $(u_{l+1}-u_{l})^4/4$.  The corresponding coupling
constant has been set to the value 1.0 to  have about the same ratio between
the average anharmonic and harmonic  potential energies as in the previous case
(approximatively 0.04 for $T=0.1$).  With such a choice, the strength of the
nonlinear terms (as measured by the  suitable perturbative parameters) are of
the same order and a comparison  between the two models makes sense. As shown
again in Fig. \ref{conduct},  the values of  $\kappa$ are now definitely
smaller (about one order of  magnitude at $N=1024$). Moreover, a linear
temperature profile sets in along  the chain and the data neatly approach a
power-law behaviour with  an exponent very close to 2/5 already for $N>64$
(diamonds in the  inset of Fig. \ref{conduct}). We therefore conclude that, 
consistently with the general argument given in the Introduction, the  cubic
nonlinearity is much less effective for what concerns the  process of energy
diffusion and transport in 1d.

In conclusion, we have shown that the dynamical correlations and transport 
in 1d lattice are strongly affected by nonlinear effects which are not 
accounted for by the simple perturbative analysis reported above even in 
the weakly anharmonic regime. For the model with cubic intersite potential, 
they manifest themselves e.g. in a faster divergence ($N^\alpha$ rather 
than $\log N$) of the thermal conductivity with the size $N$. 
Further consequences of those issues on other physical phenomena like 
energy diffusion or wavepacket propagation will be subject of future 
research.   
 
\vspace{0.2cm}
 
{\small I acknowledge useful discussions with Roberto Livi, Antonio Politi, 
Stefano Ruffo and the research group {\it Dynamics of Complex Systems} in
Florence. This work is partially supported by the INFM project 
{\it Equilibrium and nonequilibrium dynamcs in condensed matter}.} 
 
\vspace{0.5 cm}

\section*{Appendix: Evaluation of the memory function}

In this Appendix we sketch some details of the perturbative calculation of the 
memory function for the quadratic force. One wants to evaluate the Laplace 
transform of the sum appearing in Eq. (\ref{corfo3}). As a first step, we 
replace the sum for large $N$ by the integral over the Brillouin zone
\be
{1\over 2\pi} \int_{-\pi}^{+\pi} dq' \, \cos\omega(q')t \, 
\cos\omega(q-q')t \quad .
\ee
Its transform is thus given by
\bey
&&{iz\over 4\pi}\int_{-\pi}^{+\pi} \, dq' 
[ {1\over (\omega(q')+\omega(q-q'))^2 - z^2} \nonumber \\
&&+{1\over (\omega(q')-\omega(q-q'))^2 - z^2}]
\eey
It is convenient to perform the change of variable $u=e^{i(q'-q/2)/2}$. After 
some algebra, the integral reads as
\be
{z\over 2\pi} \int {du\over u}
\left[{1\over \Omega_-^2(u+1/u)^2/4 + z^2} + 
{1\over \Omega_+^2(u-1/u)^2/4 + z^2} \right]
\ee 
where the frequencies $\Omega_\pm(q)$ are defined in Eq.(\ref{omegapm})
and the integration is along a path joining the points $\pm\exp(-iq/2)$.
By means of the further change of variable $u^2=\zeta$, the integration
can be  performed on the unit circle. The result (\ref{beslap}) is thus 
obtained by means of the theorem of residues.


\begin{thebibliography}{}

\bibitem{mattis} D.C. Mattis {\it The Many-Body Problem}, (World 
Scientific, Singapore, 1993).
\bibitem{dauxois} T. Dauxois, M. Peyrard, A. R. Bishop, 
Phys. Rev. E {\bf 47} 684 (1993).
\bibitem{stillinger} F. H. Stillinger, T. Head-Gordon, C. L. 
Hirshfeld, Phys. Rev. E {\bf 48} 1469 (1993).
\bibitem{morelli} D.T. Morelli, J. Heremans, M. Sakamoto, C. Uher,
Phys. Rev. Lett. {\bf 57} 869 (1986).
\bibitem{smontara} A. Smontara, A. C. Lasjaunias, R. Maynard,
Phys. Rev. Lett. {\bf 77} 5397 (1996).
\bibitem{forsman} H. Forsman, P. Anderson, J. Chem. Phys. {\bf 80} 
2804 (1984)
\bibitem{tighe} T.S. Tighe, J.M, Worlock, M.L. Roukes, Appl. Phys.
Lett. {\bf 70} 2687 (1997).
\bibitem{rego} L.G.C. Rego, G. Kirczenow, Phys. Rev. Lett. {\bf 81} 
232 (1998).
\bibitem{leitner} D.M. Leitner, P.G. Wolynes, Phys. Rev. E {\bf 61}
2902 (2000).
\bibitem{kubo} R.Kubo, M.Toda, N. Hashitsume, {\it Statistical Physics II},
Springer Series in Solid State Sciences, Vol. 31 (1991).
\bibitem{lepri} S. Lepri, Phys. Rev. E {\bf 58} 7165 (1998).
\bibitem{pomeau} Y. Pomeau, R. R\'esibois, Phys. Rep. {\bf 19} 63 (1975).
\bibitem{noi} S. Lepri, R. Livi, A. Politi, Phys. Rev. Lett.
{\bf 78}, 1896 (1997).
\bibitem{noi2} S. Lepri, R. Livi, A. Politi, 
Europhys. Lett. {\bf 43}, 271 (1998).
\bibitem{lippi} A. Lippi, R. Livi chao-dyn/9910034 (unpublished).
\bibitem{abram} M. Abramowitz, I. Stegun (eds.), {\it 
Handbook of Mathematical Functions} (Dover, New York, 1964)
\bibitem{algor} L. Casetti, Phys. Scr. {\bf 51}, 29 (1995).
\bibitem{nose} S.~Nos\'e, J. Chem. Phys. {\bf 81}  511 (1984);
W.G.~Hoover Phys. Rev. A {\bf 31}  1695 (1985).
\bibitem{choq} Ph.~Choquard, Helvetica Physica Acta, {\bf 36}  415 (1963). 
\bibitem{noi3} S. Lepri, R. Livi, A. Politi, Physica D {\bf 119} 140(1998).
\bibitem{lieb} Z.~Rieder, J.L.~Lebowitz, E. Lieb, J. Math. Phys. {\bf 8},
1073 (1967).
\bibitem{hatano} T. Hatano, Phys. Rev. E {\bf 59} R1 (1999)

\end{thebibliography}
\end{document}